\documentclass[9pt,twocolumn,twoside]{opticajnl}
\journal{opticajournal} % use for journal or Optica Open submissions

\setboolean{shortarticle}{true}
\usepackage{lineno}
\title{Three-dimensional abruptly autofocusing by counter-propagating Airy pulses with radial Airy beam profile}

\author[1]{Youngbin Park}
\author[2]{Xiaolin Su}
\author[2,*]{Qian Cao}
\author[1,3,*]{Andy Chong}

\affil[1]{Department of Physics,Pusan National University, Geumjeong-Gu, Busan, 46231, Republic of Korea}
\affil[2]{School of Optical-Electrical and Computer Engineering, University of Shanghai for Science and Technology, Shanghai, 200093, China}
\affil[3]{Institute for Future Earth, Pusan National University, Geumjeong-Gu, Busan 46241, Republic of Korea}

\affil[*]{chong0422@pusan.ac.kr and cao.qian@usst.edu.cn}

\begin{abstract}
We report the experimental observation of a three-dimensional abruptly autofocusing effect by synthesizing a radially distributed Airy beam with two counter-propagating Airy pulses in time. As the wave packet propagates in a dispersive medium, the radially distributed Airy beam converges inward to the center point. Two Airy pulses counter-propagate toward each other to merge to form a high peak power pulse. As the result, the high intensity emerges abruptly as the wave packet achieves three-dimensional focusing. This autofocusing effect is believed to have potential applications such as material modification, plasma physics, nanoparticle manipulations, etc.
\end{abstract}

\setboolean{displaycopyright}{false} % Do not include copyright or licensing information in submission.

\begin{document}

\maketitle

\section{Introduction}

Diffraction and dispersion, which tend to distort optical wave packets during propagation, are ubiquitous phenomena. However, certain optical waves, known as localized waves, maintain their shape during propagation. Certain localized beams such as Bessel beams, Airy beams, etc. are well known for resisting diffraction \cite{saleh:08}. These localized beams exhibit a unique property of propagation invariance so-called non-diffractive propagation. Another unique property is self-healing effect, where localized beams reconstruct their spatial shape during propagation even though parts of the beam are blocked \cite{Shen:22}. With these properties, the localized beams have numerous potential applications, such as micromanufacturing \cite{Amako:03}, tomography \cite{Lorenser:14} and optical communication \cite{Zhao:19}. 

In contrast to two-dimensional beams, the Airy beam is a one-dimensional localized beam as the solution of a paraxial equation \cite{Berry:79}. As the localized beam, the Airy beam also exhibits non-diffractiveness during propagation along with the self-healing poroperty \cite{Broky:08}. However, the peak of the Airy beam follows a unique parabolic trajectory during propagation, a phenomenon known as free-acceleration \cite{Siviloglou:07}. With these properties, the Airy beam has found numerous applications, including nonlinear optics \cite{ellenbogen:09,Dolev:10}, plasma physics \cite{Polynkin:09}, optical communications \cite{Hu:18}, metasurface \cite{Li:15}, and high energy physics \cite{Campos:24}. There are various methods to generate Airy beams, but a common approach is to use a spatal light moudulator (SLM) \cite{Siviloglou:077}.

The free-acceleration effect enables a unique beam manipulation. When the Airy beam profile is radially distributed (in the cylindrical oordinates), the beam free accelerates toward the center and eventually forms a highly intense central spot. This process is called autofocusing since the focusing effect occurs solely through free-space propagation. At the same time, the high-intensity spot emerges abruptly, transitioning from the nearly zero intensity to the high intensity. This abruptly autofocusing effect was first reported, being characterized by two-dimensional radially symmetric Airy beams converging to the center point during propagation \cite{Papazoglou:11}. This abruptly autofocusing phenomenon is widely exploited in various applications \cite{Papazoglou:13,Zhang:11,Courvoisier:16} and also observed in other beams, such as vortex beams and Pearcey beams \cite{Davis:12, Ring:12}.

In time, a pulse is broadened due to the group velocity dispersion (GVD) effect. However, since the pulse propagation is also governed by a one-dimensional paraxial equation, an Airy pulse remains non-dispersive during propagation. In addition to its non-dispersive propagation, an Airy pulse also exhibits self-healing and self-acceleration in the presence of GVD \cite{Chong10}. By placing two counter-propagating Airy pulses, they accelerate toward each other in a dispersive material and eventually constructively interfere, resulting a high-peak-power pulse. This process can be understood as the temporal autofocusing.

By combining the temporal autofocusing and the spatial autofocusing of the
radial Airy beam, a three-dimensional(3D) abruptly autofocusing wave packet is predicted \cite{Papazoglou:11}. In this manuscript, we demonstrate this 3D abruptly autofocusing wave packet. First, two counter-propagating Airy pulses are generated using an SLM based pulse shaper \cite{Weiner00}. The counter-propagating Airy pulses acquire the radially symmetric Airy beam profile by using another SLM. The spatiotemporal profile of this abruptly autofocusing wave packet is measured by using a 3D pulse intensity diagnostic technique \cite{Li:11}. This wave packet has potential applications in many fields such as nanomaterial processing.

%\author{Author One\authormark{1} and Author Two\authormark{2,*}}

%\address{\authormark{1}Peer Review, Publications Department,
%Optica Publishing Group, 2010 Massachusetts Avenue NW,
%Washington, DC 20036, USA\\
%\authormark{2}Publications Department, Optica Publishing Group,
%2010 Massachusetts Avenue NW, Washington, DC 20036, USA\\
%%\authormark{3}xyz@optica.org}

%\email{\authormark{*}xyz@optica.org}}

%Example with the corresponding author designated by an asterisk and a note indicating equal contributions by two authors.

%\author{Author One\authormark{1,3} and Author %Two\authormark{2,3,*}}

%\address{\authormark{1}Peer Review, Publications Department,
%Optica Publishing Group, 2010 Massachusetts Avenue NW, %Washington, DC 20036, USA\\
%\authormark{2}Publications Department, Optica Publishing Group, %2010 Massachusetts Avenue NW, Washington, DC 20036, USA\\
%\authormark{3}The authors contributed equally to this work.\\
%\authormark{*}xyz@optica.org}}

%\section{Examples of Article Components}
%\label{sec:examples}

\section{Theory and numerical Simulation}

Spatiotemporal wave packets in a dispersive medium are governed by a 3D paraxial wave equation:
\begin{equation}
\frac{\partial^2 \psi}{\partial x^2} + \frac{\partial^2 \psi}{\partial y^2} + 2ik \frac{\partial \psi}{\partial z} 
- \frac{k \beta_2}{2} \frac{\partial^2 \psi}{\partial T^2} = 0
\end{equation}

\vspace{0.5em} 
\noindent Here,  \(k\) is the wave vector of light and \(z\) is the propagation distance. The parameter \(\beta_2\) denotes a GVD coefficient, defined as \(\beta_2 = \partial^2 k / \partial \omega^2\)and \(T = t - (z / v_g)\) represents the local time frame, where \(v_g\) is the group velocity of light. For the counter-propagating two Airy pulses, which will be referred to as dual Airy pulses, with the radially distributed Airy beam, which will be referred to as the Airy ring beam, in this manuscript, the wave packet is mathematically described as,
\begin{figure}[b]
    \centering
    \includegraphics[width=0.45\textwidth,keepaspectratio]{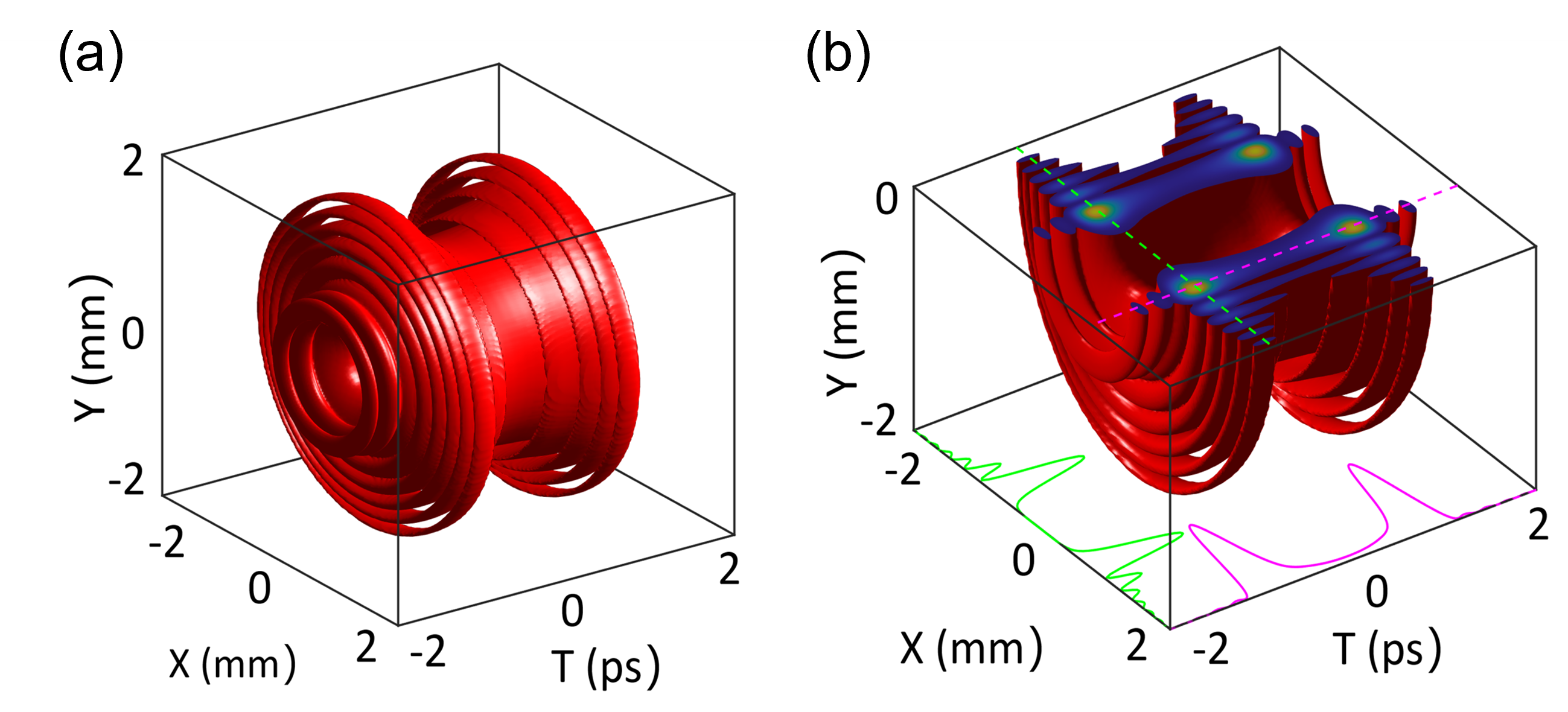} 
    \caption{Numerical result of the generation 3D Airy ring beam - dual Airy pulse wave packet. (a) Three-dimensional(3D) iso-intensity profile of the wave packets (b) 2D cross-section of the intensity profile of the wave packets}
    \label{fig1}
\end{figure}
\begin{figure}[t]
    \centering
    \includegraphics[width=0.45\textwidth,keepaspectratio]{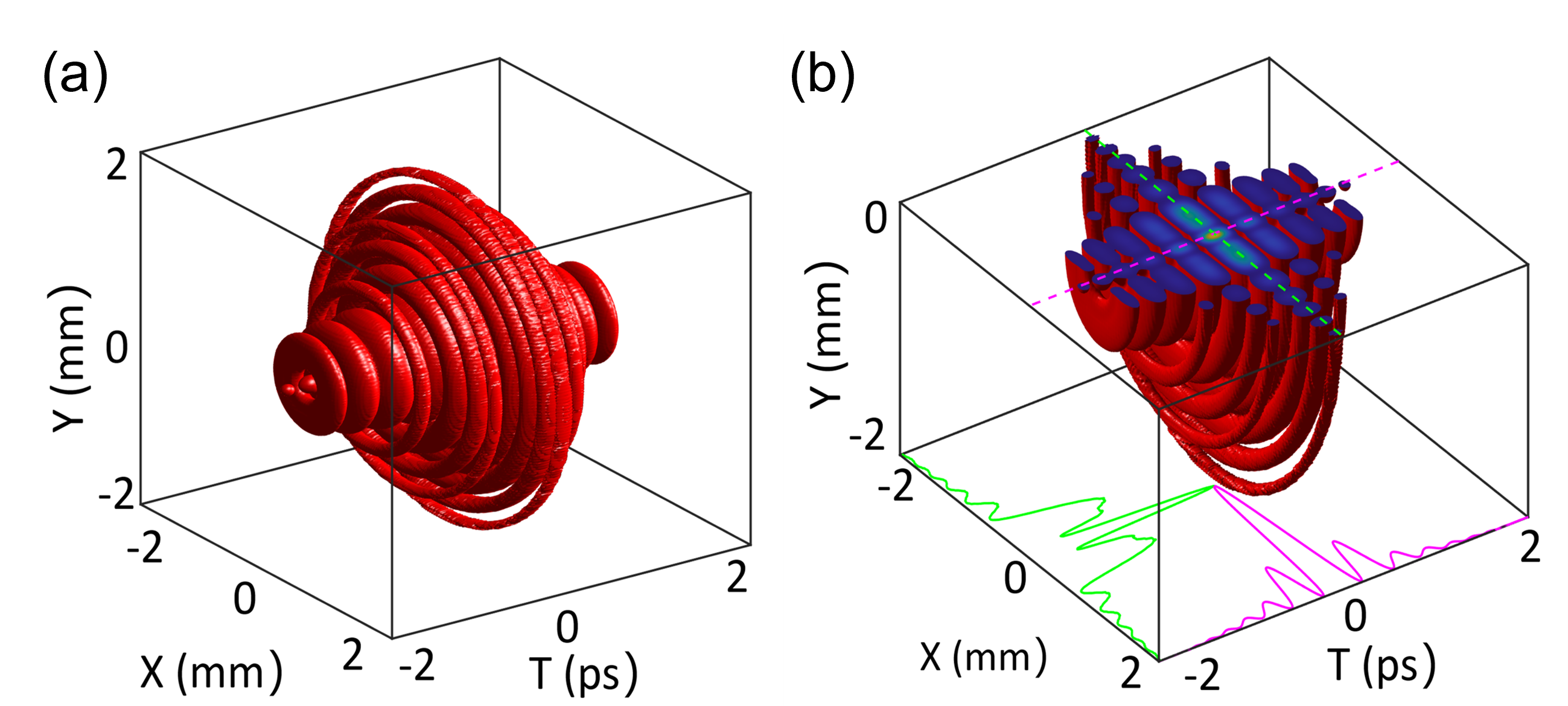} 
    \caption{Numerical simulation of the Airy ring-Airy pulses wave  packets after the propagation. (a) Iso-intensity of the autofocused wave packets (b) The inside view of the wave packets and their intensity distribution with the Autofocusing effect}
    \label{fig2}
\end{figure}
\begin{multline}
\psi\left(r,T\right) = A_0 \, \text{Ai}\left(\frac{r-r_0 }{w}\right) 
\exp\left\{\alpha \left(r_0 - r\right)\right\} \\ 
\times \Bigg[ \text{Ai}\left(\frac{T- b}{T_0}\right)\exp\left\{\gamma \left(T - b\right)\right\} \\  
+ \text{Ai}\left(-\frac{T+ b}{T_0}\right)\exp\left\{-\gamma \left(T + b\right)\right\}
\Bigg]
\end{multline}

\noindent In Eq. (2), \(A_0\) is the amplitude of the wave packets, Ai$\left(x\right)$ is an Airy function, \(r_0\) is the radius of an initial Airy ring beam and \(r\) is the cylindrical radial distance where \(r = \sqrt{x^2 + y^2}\). \(w\) is a scaling factor to determine the radial width of the Airy ring beam, as \(\alpha\) denotes the decaying parameter to generate a finite-energy Airy ring beam. In time, \(\gamma\) denotes the decaying factor for finite-energy Airy pulses, \(T_0\) determines the temporal width of wave packets, and \(b\) sets the temporal location of the initial Airy pulses. Eq. (2) represents the superposition of two time-reversed Airy pulses, each possessing the same radial Airy ring beam distribution. 

A numerical simulation has been performed to demonstrate the 3D autofocusing effect of the suggested optical wave packet. Fig \ref{fig1} visualizes the initial iso-intensity profile and the internal intensity distribution of the wave packet. As the wave packet propagates through a dispersive medium, the radial Airy ring converges to the center, while the Airy pulses merge in the middle as shown Fig \ref{fig2} clearly illustrates the 3D abruptly autofocusing effect.

\section{Experimental Results}
An initial pulse is emitted from a mode-locked ytterbium(Yb) fiber oscillator. The oscillator has a Gaussian beam output with a mode field diameter(MFD) of approximately 1 mm. The laser spectral bandwidth is \textasciitilde80nm at the 1030nm, which corresponds to a Fourier transform limited pulse duration of approximately 54fs.
\begin{figure}[b]
    \centering
    \includegraphics[width=0.45\textwidth,keepaspectratio]{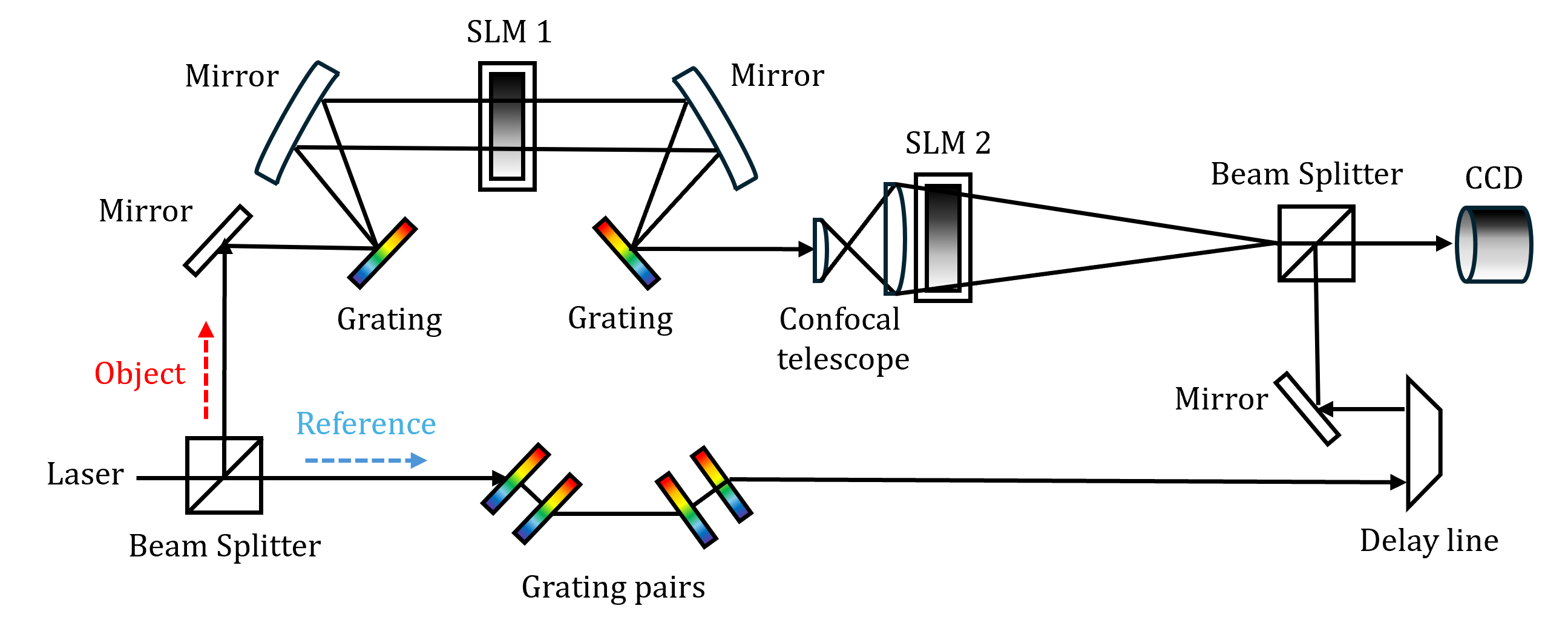} 
    \caption{Schematic of an experimental setup}
    \label{fig3}
\end{figure}
\noindent We describe the experimental method to generate the wave packets in Fig \ref{fig3}. The initial pulsed beam is split into two paths. The object beam is modulated using a SLM based pulse shaper \cite{Weiner00}. To convert the initial pulsed Gaussian beam into two counter propagating Airy pulses, we utilize the method described in ref \cite{Wan:14}. In detail, linear phases with an opposite slope (i.e. triangular spectral phase) are imposed on the first SLM(SLM 1), resulting two temporally seperated pulses. Subsequently, a cubic spectral phase is applied on each of these pulse, transforming them into Airy pulses. By controlling the sign of the cubic spectral phase, the direction of the tail for each Airy pulse can be adjusted to form dual Airy pulses that face each other. Such phase pattern is described as Eq. (3) and shown in Fig \ref{fig4}(a).
\begin{multline}
\varphi_{\text{SLM 1}} = a_0 \left(\xi + \xi_0\right) - a_0 \left(\xi - \xi_0 \right) \\
- a_1 \left(\xi + \xi_1 \right)^3 + a_1 \left(\xi - \xi_1 \right)^3
\end{multline}
\noindent In the pulse shaping system, the input pulse is Fourier-transformed from the time domain to the frequency domain on the SLM 1. Then, we can modulate the spectral phase by adjusting the phase on the SLM 1 according to Eq. (3), where \(\xi\) represents a horizontal coordinate on the SLM 1. The first and second terms in Eq. (3) are responsible for converting the initial pulse into two separate pulses. The third and fourth represent the phase modulation to shape each pulse into an Airy pulse.

Following pulse shaping, the beam is expanded three times using a telescope. The second SLM(SLM 2) works as a beam shaper, modifying the Fourier spatial frequency components, according to the Fourier transform in the spatial domain. The phase applied on the SLM 2 is given by Eq. (4) (shown in Fig \ref{fig4}(b)).

\begin{equation}
    \varphi_{\text{SLM 2}} = -b_0\left(\rho-\rho_1\right) - b_1 \left( \rho - \rho_2 \right)^3
\end{equation}

\noindent In Eq. (4), \(\rho = \sqrt{\xi^2 + \eta^2}\) is a radial component in polar coordinates, where \(\eta\) is the vertical coordinate on the SLM 2. The first phase term transforms the input beam into a ring-shaped beam, while the second converts it into an Airy ring beam. 

Applying a conic spatial phase, similar to the pulse shaping mechanism, generates the ring beam. Introducing a cubic phase in the radial direction induces a radial Airy profile onto the spatial ring beam. As a result, an Airy ring beam - dual Airy pulse wave packet is formed. Theoretically, the transformed beam shape can be expressed as the Hankel transform of a Gaussian beam as shown in Eq. (5):
\begin{figure}[t]
    \centering
    \includegraphics[width=0.45\textwidth,keepaspectratio]{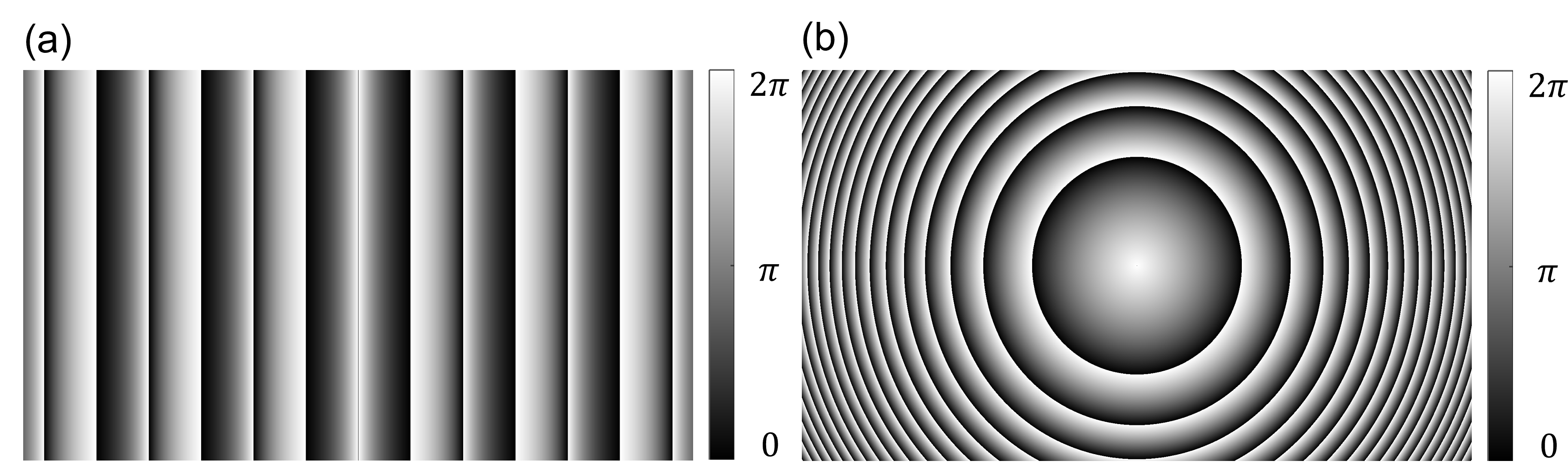} 
    \caption{Phase patterns for generating (a) dual Airy pulse in SLM 1 and (b) Airy ring beam in SLM 2.}
    \label{fig4}
\end{figure}

\begin{figure}[b]
    \centering
    \includegraphics[width=0.45\textwidth,keepaspectratio]{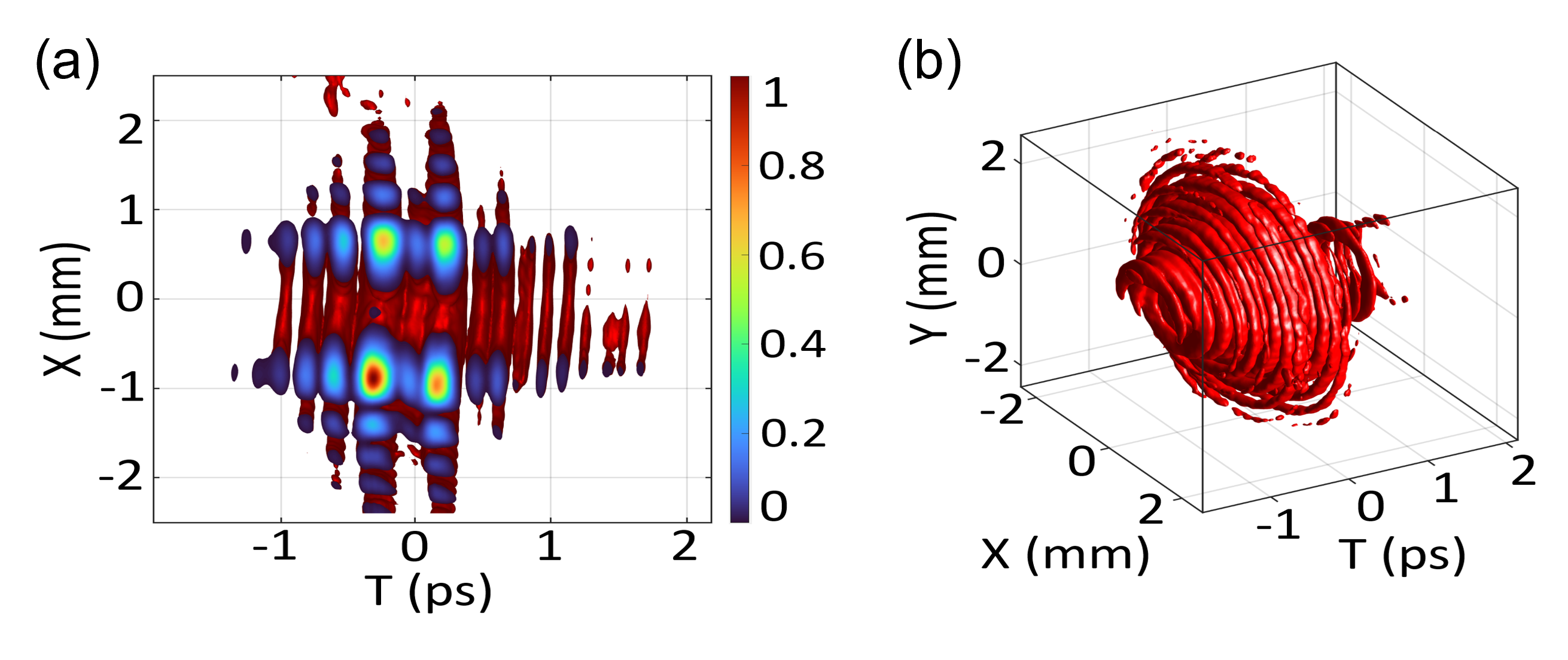} 
    \caption{Experimental results of the generation 3D Airy ring beam - dual Airy pulse wave packet. (a) Two-dimensional(2D) cross-section of the intensity profile of the wave packets and (b) 3D iso-intensity profile}
    \label{fig5}
\end{figure}

\begin{equation}
G\left(r\right)=\int_{0}^{\infty} \rho g\left(\rho\right) J_0\left(2\pi\rho r\right) \exp{\{i\varphi_{\text{SLM 2}}\left(\rho\right)\}} \, d\rho    
\end{equation}

\noindent Here, \(g(\rho)\) represents the initial beam profile, assumed to be a Gaussian, and \(J_0\) denotes the Bessel function of the first kind of order zero, For a small beam sizes, the \(J_0(x)\) can be approximated using the Taylor expanded as $J_0\left(x\right)\approx1-\frac{1}{2}x^2\approx\cos(x)$. Hence, the Hankel transform can  be approximated as a cosine Fourier transform of a radial Hermite-Gaussian function. In the range of (\(\rho > 0\)) the Hermite-Gaussian function closely resembles a shifted Gaussian function. As a result, the transformed beam can be approximated as a Fourier transform of a shifted Gaussian function with a cubic phase ultimately forms an Airy function in the radial direction.

On the other optical path, the pulse is dechirped using a grating pair to produce the transform-limited(reference) pulse. The 3D intensity measurement is performed by overlapping the object and reference wave packets utilizing the diagnostic technique described in \cite{Li:11}. As the object and reference wave packets are combined at a slight tilt angle, an interference pattern is generated. Since the reference pulse is substantially shorter, the interference pattern reveals the intensity distribution of the object wave packet at the reference pulse temporal position. This process is repeated to capture numerous interference patterns and therefore the intensity distributions at a various reference pulse delays. By combining these intensity distributions, the 3D intensity profile of the wave packet can be reconstructed.

\begin{figure}[t]
    \centering
    \includegraphics[width=0.45\textwidth,keepaspectratio]{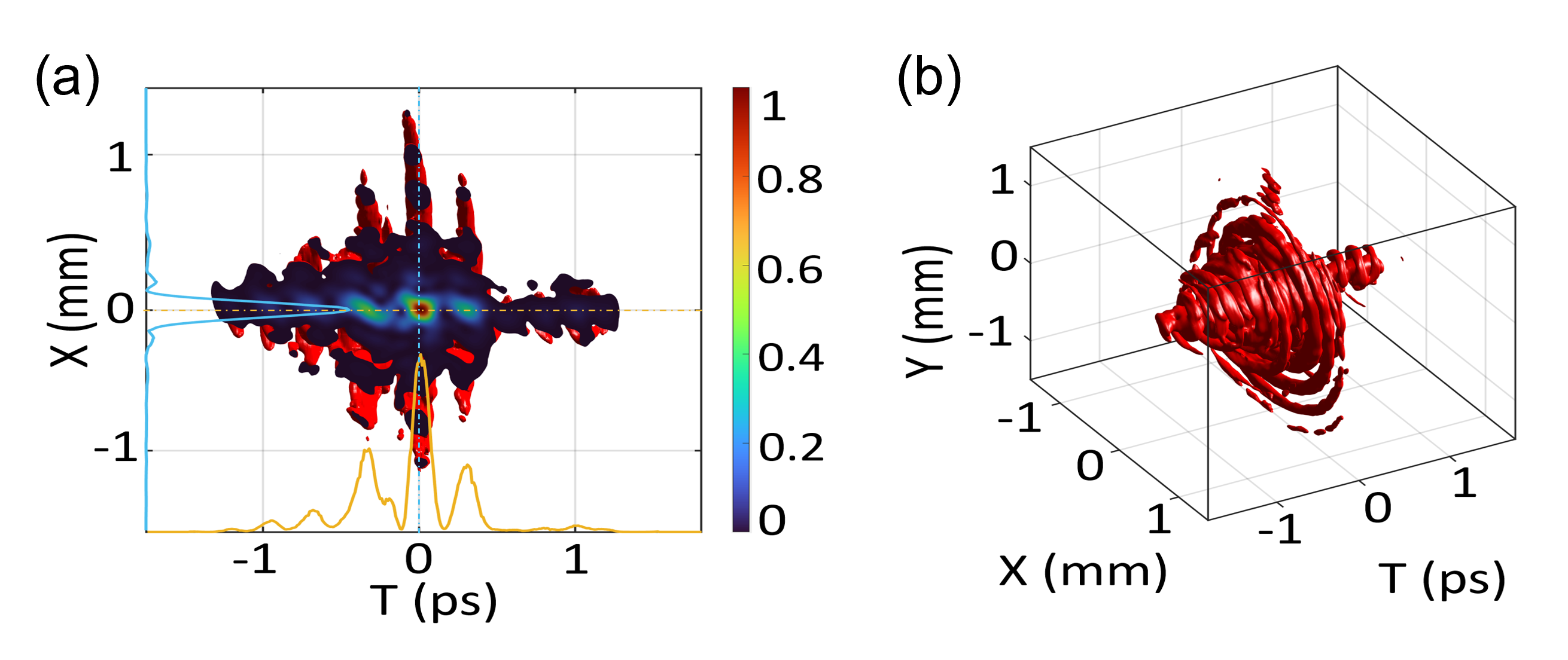} 
    \caption{Experimental results of the Airy ring-Airy pulses wave packets after the propagation. (a) 2D cross-section of the intensity profile of the wave packets and (b) 3D iso-intensity profile}
    \label{fig6}
\end{figure} 

The experimentally generated Airy ring beam - dual Airy pulse wave packet is shown in Fig \ref{fig5}. Fig \ref{fig5}(a) denotes the cross-section of the wave packets while the 3D iso-intensity profile of the wave packet is shown in Fig \ref{fig5}(b). The spatial profile reveals the ring beam pattern with tails extending in radial direction, as predicted by the Airy ring beam structure. In the temporal domain, the cross-section of the wave packets clearly show two pulse structures with tails extending in opposite directions, as predicted by the dual Airy pulse configuration. The Airy ring beam - dual Airy pulse wave packet propagates through a 4-inch-long SF11 glass. This material induces enough diffraction and dispersion effect, causing the wave packet to autofocus in both space and time. The measured autofocused wave packet is shown in Fig \ref{fig6}(a) and (b). In the experiment, an initial Airy ring beam with a diameter of \textasciitilde1.8 mm and is autofocused to a beam size of \textasciitilde130 \(\mu\)m.  Meanwhile, Airy pulses separated by \textasciitilde500 fs are temporally autofocused to a pulse with a duration of \textasciitilde110 fs, due to propagation through the SF11 glass. The measurement clearly shows a high intensity spot appearing at the center of the wave packet, providing evidence of the abruptly autofocusing phenomenon.

\section{Conclusion}
We have demonstrated the 3D abruptly autofocusing phenomenon by the Airy ring beam- dual Airy pulse wave packet. As this wave packet propagate through the dispersive material, a high-intensity spot forms at the center, providing evidence of the abruptly autofocusing phenomenon. This wave packets holds potential for a variety of applications, such as micromachining and other precision material processing.

% Bibliography
\bibliography{sample}

% Full bibliography added automatically for Optics Letters submissions; the following line will simply be ignored if submitting to other journals.
% Note that this extra page will not count against page length
\bibliographyfullrefs{sample}

%Manual citation list
%\begin{thebibliography}{1}
%\bibitem{Zhang:14}
%Y.~Zhang, S.~Qiao, L.~Sun, Q.~W. Shi, W.~Huang, %L.~Li, and Z.~Yang,
 % \enquote{Photoinduced active terahertz metamaterials with nanostructured
  %vanadium dioxide film deposited by sol-gel method,} Opt. Express \textbf{22},
  %11070--11078 (2014).
%\end{thebibliography}

% Please include bios and photos of all authors for aop articles
\ifthenelse{\equal{\journalref}{aop}}{%
\section*{Author Biographies}
\begingroup
\setlength\intextsep{0pt}
\begin{minipage}[t][6.3cm][t]{1.0\textwidth} % Adjust height [6.3cm] as required for separation of bio photos.
  \begin{wrapfigure}{L}{0.25\textwidth}
    \includegraphics[width=0.25\textwidth]{john_smith.eps}
  \end{wrapfigure}
  \noindent
  {\bfseries John Smith} received his BSc (Mathematics) in 2000 from The University of Maryland. His research interests include lasers and optics.
\end{minipage}
\begin{minipage}{1.0\textwidth}
  \begin{wrapfigure}{L}{0.25\textwidth}
    \includegraphics[width=0.25\textwidth]{alice_smith.eps}
  \end{wrapfigure}
  \noindent
  {\bfseries Alice Smith} also received her BSc (Mathematics) in 2000 from The University of Maryland. Her research interests also include lasers and optics.
\end{minipage}
\endgroup
}{}

\end{document}